# Review of sound card photogates

**Zoltán Gingl[1], Róbert Mingesz[1], Péter Makra[2] and János Mellár[1]**

[1]Department of Technical Informatics, University of Szeged, Árpád tér 2, 6720 Szeged, Hungary

[2]Department of Experimental Physics, University of Szeged, Dóm tér 9, 6720 Szeged, Hungary

e-mail: gingl@inf.u-szeged.hu

**Abstract.** Photogates are probably the most commonly used electronic instruments to aid experiments in the field of mechanics. Although they are offered by many manufacturers, they can be too expensive to be widely used in all classrooms, in multiple experiments or even at a home experimentation. Today all computers have a sound card – an interface for analogue signals. It is possible to make very simple yet highly accurate photogates for cents, while much more sophisticated solutions are also available at a still very low cost. In our review we show several experimentally tested ways of implementing sound card photogates in detail, and we also provide a full-featured, free, open-source photogate software as a much more efficient experimentation tool than the usually used sound recording programs. Further information is provided in a dedicated page, www.noise.physx.u-szeged.hu/edudev.

## 1. Introduction

Photogates are probably the most frequently used motion detectors in demonstration experiments of mechanical movements. Photogates can be used to detect the time of various events quite precisely, they are easy-to-use, they employ a sufficiently simple principle that allows students to easily focus on the investigated phenomena. Although there are instruments for the direct measurement of displacement, velocity and acceleration, the lower cost and simplicity of photogates makes them more popular at several teaching levels.

Professional photogates can still be expensive, which is especially problematic if the teacher would like to involve many students in experimenting at several lab sites simultaneously or if the teacher would like to give an experimental homework. The typical solution to lower the costs dramatically is based on capturing real-time signals with multimedia computers [1-7]. Today these computers are widely available and all of them have input and output for analogue signals via the audio interface, the so-called sound card. This port can sample a time-dependent voltage at rates up to (and even beyond) 44 kHz, which yields excellent resolution in time, meeting the requirements of mechanical experiments. Although the absolute accuracy of the measured amplitude is unreliable and poor, the resolution, relative accuracy and linearity are good, and what is more important, the timing is precise, which is ideal for photogates.

As the large number of web pages and journal articles attest, many teachers, experimenters, hobbyists and even scientists use sound card photodetectors and photogates [8-12]. The most common solution is to use a separate light source, a photodetector connected to the audio input with a few additional components and apply a free sound recording software to visualise the shadings caused by a moving object. Although it works and the cost is very low, apart from being questionable didactically, this method has additional shortcomings: due to the properties of the audio input, only changes can be observed - there is no information about the actual value of the signal. Furthermore, the use of software is inconvenient; additional time is needed to extract the time information, and in general, it is not handy at all. However, there are solutions available to overcome these problems, so sound card photogates can even compete with professional devices. In this paper we review most of the hardware possibilities and provide important addition with software especially developed for photogate applications. Solutions for the problems associated with sound card properties, the absolutely simplest, lowest cost hardware and more sophisticated realisations are all exposed.

## 2. The anatomy of the photogate

The operating principle of the photogate is rather simple: the lightbeam of an emitter is directed to the photodetector, therefore if a moving object gets in between, it blocks the light from the detector. The detector has more or less a two-state output, so the time of the beginning and ending of light interruption can be measured with level crossing detection. The light source is intensive enough to minimise the effect of ambient light, and in most cases optical shielding, filtering and infrared emitter-detector pairs are used to further reduce sensitivity to parasitic components. Figure 1 shows the typical transmissive (also known as interrupter) and reflective operation mode arrangements.

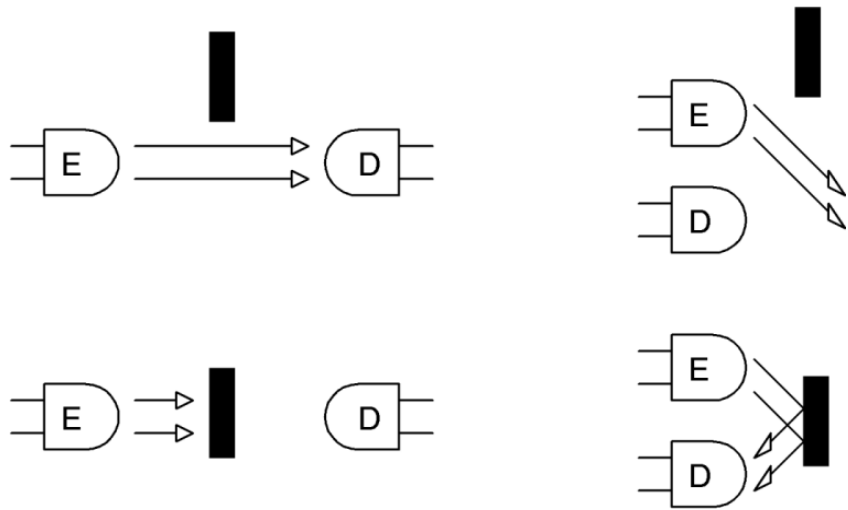

**Figure 1.** The principle of transmissive (aka interrupter) and reflective photogates. The transmissive is the most frequently used mode, whereas the reflective mode is more sensitive to the shape and surface properties of the object, and the reflected signal is significantly smaller.

### 2.1. Light sources

Best light sources provide small-diameter and intensive light beam to allow clear detection of even small objects. It is desirable to have narrow optical bandwidth and of course it should be matched to the sensitivity range of the detector, or at least the ranges must overlap quite well. Table 1 shows the light sources most frequently applied in sound card photogates.

**Table 1.** Light sources

| Light source | Advantages | Disadvantages |
|---|---|---|
| Ambient light | no cost | uncertain,<br>weak,<br>spatially distributed |
| Mains lamp | sufficiently strong signal<br>stable light source | component at 2×mains frequency (50/60Hz)<br>bulky<br>spatially distributed |
| DC supply lamp | sufficiently strong signal<br>stable light source | bulky<br>spatially distributed |
| Flashlight | sufficiently strong signal<br>stable light source | battery may be accidentally discharged |
| USB lamp | sufficiently strong signal<br>stable light source<br>no external power needed | the cable may be too short<br>bulky<br>spatially distributed |
| Laser pointer | strong signal<br>small diameter<br>small divergence lightbeam<br>very narrow optical bandwidth | battery may be accidentally discharged;<br>needs precise mounting |
| Infrared LED | sufficiently strong lightbeam<br>narrow optical bandwidth | needs power source<br>spatially distributed |

## *2.2. Photodetectors*

The photodetector should be sensitive, fast, and have small sensing surface to provide the best spatial resolution. The following table summarises the properties of the most commonly used detectors. The detectors and their most important properties are listed in table 2. Professional photogates (and many remote control devices) employ infrared LED-phototransistor pairs, due to their better selectivity and reduced sensitivity to unwanted light sources. This is the optimal choice; however, other combinations can also be acceptable.

**Table 2.** Light detectors

| Detector | Advantages | Disadvantages |
|---|---|---|
| Phototransistor 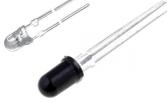 | sensitive<br>fast response<br>small sensing surface<br>wide selection<br>lowest cost | bias voltage needed |
| Photoresistor (photocell) 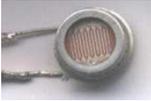 | sensitive<br>low cost | bias voltage needed<br>slow rise and fall times<br>larger surface |
| Photodiode 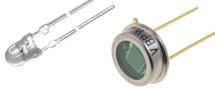 | fast<br>small sensing surface<br>no bias required<br>wide selection | small output signal<br>higher cost |
| solar cell 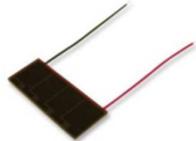 | no bias required<br>voltage output | small output signal<br>very large surface |

## 2.3. Event detection, time measurement

The event detection and time measurement unit makes the photogate complete. This part detects the transitions of the output signals and measures the time between these events using a periodically driven counter.

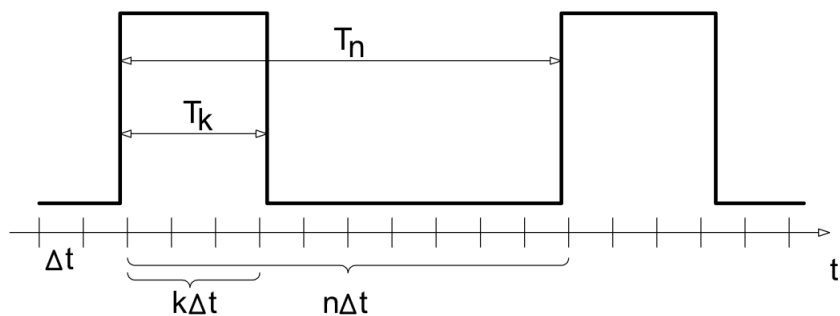

**Figure 2.** The time $T_n$ between mechanical events associated with light interruptions detected by the photogate can be measured by counting the time units $\Delta t$ between the subsequent events, so $T_n \approx n\Delta t$. $T_k \approx k\Delta t$ measures the time of light interruption, often used to calculate instantaneous object speed.

Depending on the set of features required, we can use the unit to register just this value, or apply a microprocessor to do further processing and provide a more sophisticated display.

If the raw output signal of the photodetector is digitised and transmitted into a personal computer, most of the signal processing can be done by software - this is the most flexible solution.

## 3. How to interface photogates to a PC

Personal computers have some ports to connect to external signals and devices. Most of these ports are digital and require special communication protocols. On modern computers USB is the most commonly used general-purpose port to connect to external devices, but it requires external hardware and driver software as well.

The sound card is the only port for analogue input-output signals. Only alternating voltages can be measured and generated in the range of about 1 $V_{rms}$. The photodetectors listed above can easily be arranged to provide voltage output with the help of one or two additional electronic components. Many light sources are completely external, but some can be driven from the computer, as will be shown later.

### 3.1. Detector interfacing possibilities

Phototransistors are often used in the so called common emitter configuration to give light dependent output voltage. The emitter is grounded, the collector is connected to a positive voltage source via a series resistor. The higher the light intensity - when higher current is flowing through the transistor - the lower the collector voltage in this arrangement. A very similar method works for photoresistors, when a voltage divider can be built using the serially connected resistor and photoresistor biased with a voltage source. In both cases the detectors work like light controlled switches.

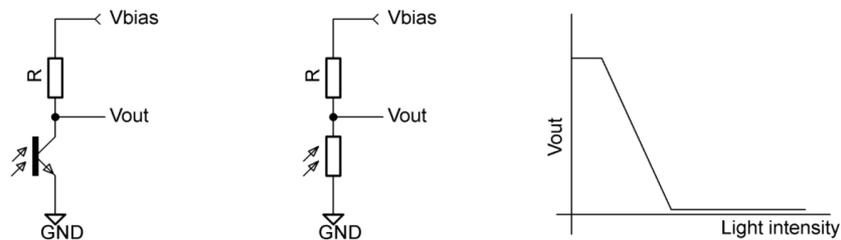

**Figure 3.** Phototransistor in a common emitter configuration and photoresistor in a voltage divider arrangement can be considered as a light controlled switch.

Photodiodes can be operated in photovoltaic (voltage output) or photoconductive (current output) mode; in the latter case a parallel termination resistor is used. Solar cells are practically large area photodiodes operated in photovoltaic mode, therefore the output signal can be considered as voltage.

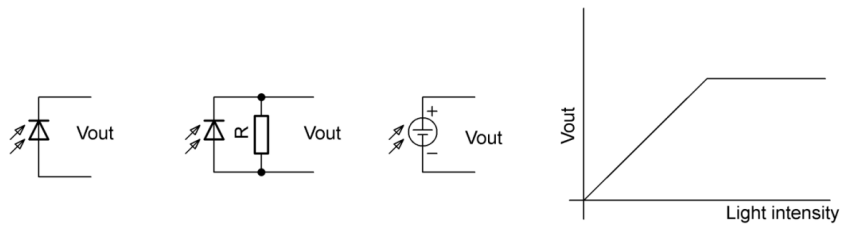

**Figure 4.** Photodiodes can be operated in photovoltaic or photocurrent mode (terminated with a resistor). Solar cells can be thought of as large area photodiodes working in photovoltaic mode.

### 3.1.1. Internally biased photodetectors

The sound card has two kinds of input ports for external devices: microphone input and line input. Both accept voltage in the range of about 1 $V_{rms}$, but the microphone port may have a "boost" (gain of 10 or so) option to accept even smaller signals. In most cases PC microphones are electret microphones that need power, therefore to eliminate the necessity for batteries, the microphone input provides the required bias voltage according to the AC'97 specification (for an example, see ADAU1961 codec datasheet [13]). Figure 5 shows the possible microphone input configurations.

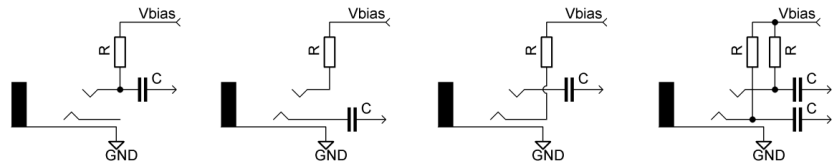

**Figure 5.** Simplified microphone input schematics. The left configuration is the most frequent; on the right hand side the true stereo input is shown.

This means that all photodetectors can be directly connected to the microphone input port of a PC, no additional components are needed [11]. Since cheap photodetectors may be purchased for cents, extremely low-cost sound card photogates can be made.

One can easily measure $V_{bias}$ by connecting a voltmeter between the tip or ring and the sleeve of the plugged in 3.5 mm jack plug. Replacing the voltmeter with an ammeter will show the current $V_{bias}/R$.

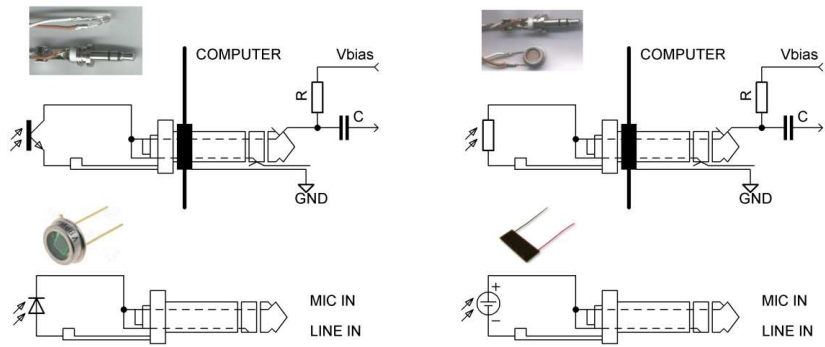

**Figure 6.** Connecting photodetectors to the audio inputs. The phototransistors and photoresistors need bias voltage, therefore they can be directly connected only to the microphone input. Photodiodes and solar cells can be connected to the line input as well, which is the preferred solution (see section "Microphone versus line audio input ports")

### 3.1.2. Externally biased photodetectors

Photodetectors can be connected to the line input as well if an external bias voltage is provided. Batteries are commonly used, but the USB port also has a 5 V supply terminal (pin 1, the wire colour is red), therefore one can eliminate the risk of discharged batteries.

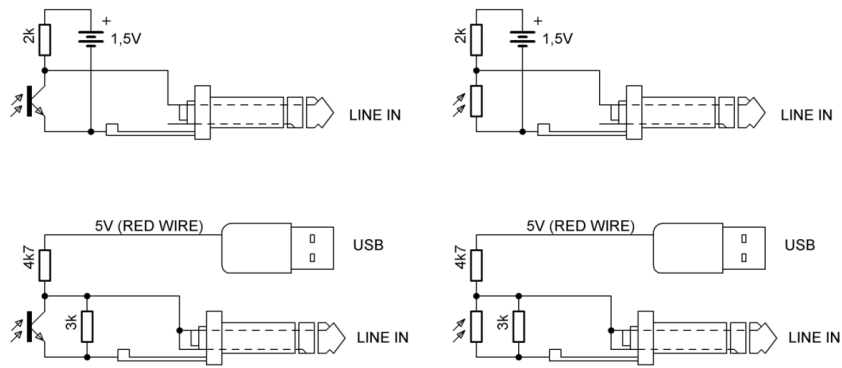

**Figure 7.** Externally biased phototransistor and photoresistor. Series resistors are included to limit voltage excursions to about 2 V and thus avoid overdriving the audio inputs.

### 3.2. Protecting the audio inputs

The internal analogue circuitry is typically powered from 5 V, therefore the safe internal signal range is 0 V–5 V. The inputs are connected to the internal circuitry via a capacitor, where the average voltage is set close to half of the supply, 2.5 V. This means that if the input signal abruptly changes by $\Delta V$, the same change will be observed at this internal point. If $|\Delta V|<2.5$ V, the signal will always stay in a safe range. This is ensured if the bias voltage used for the photodetector is less than 2.5 V (either from the microphone port or from a battery). Figure 8 shows a measurement on a simple circuit similar to the microphone input of the sound card. The component values correspond to a typical audio input. The input and output signals of the circuit realised on a prototype board were captured by a Rigol VS5042D oscilloscope; the light beam was abruptly interrupted and uninterrupted to generate the typical waveform at the collector of the phototransistor.

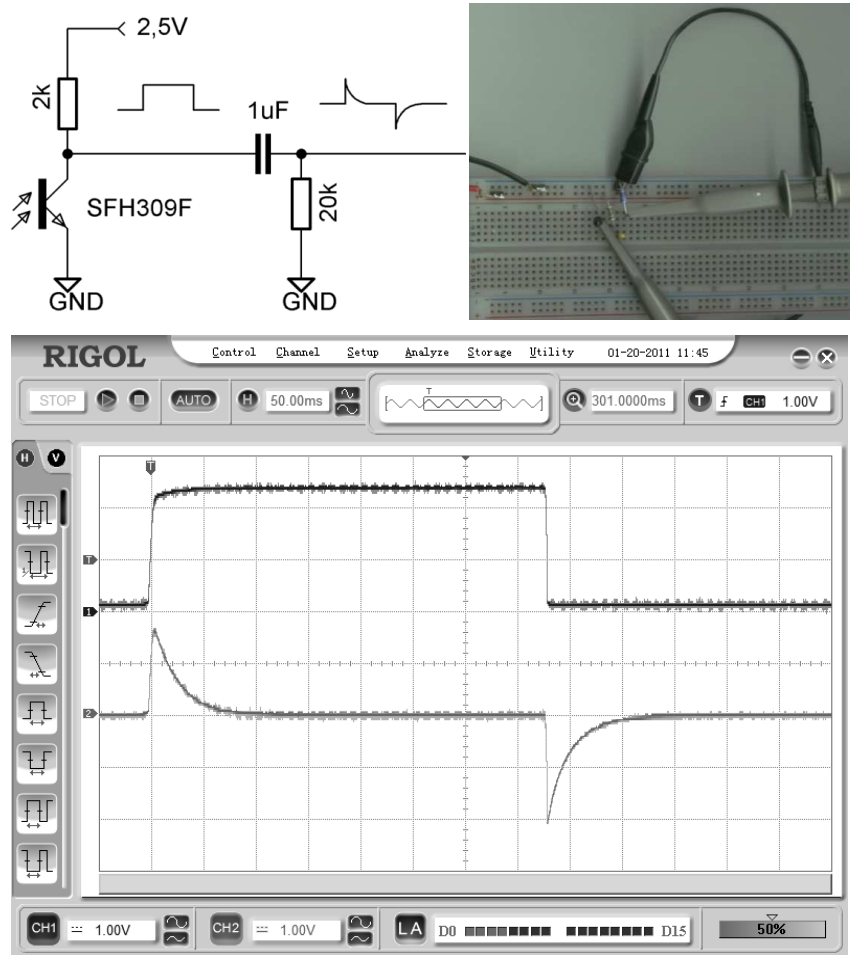

**Figure 8.** Measured input and output waveforms for an AC coupling stage.

If a larger voltage swing is possible at the input, it can be connected to the audio input via a series resistor that will limit the maximum possible inrush current below the safe limit, typically 10 mA (see for example the ADAU1961 codec absolute maximum ratings [13]). If the bias voltage is $V_{bias}$, the resistor should be $V_{bias}/I_{max}=V_{bias}/0.01A$.

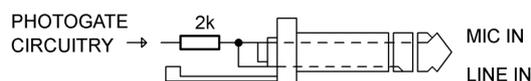

**Figure 9.** Connecting the photodetectors to the input via a series resistor limits the inrush current transients into a safe range. Resistance value of 2k ensures protection for voltage transients as high as 10 V.

Note that the photodetectors already limit the current into a safe range in most cases, it is very unlikely that they can carry such a high current that can cause damage to the internal circuitry. Anyway, using a series resistor of 1k-2k provides reliable protection in all cases even if the bias voltage is provided by a 9V battery.

It is rarely mentioned in sound card projects that user's safety and protection of the computer's internal circuitry from electrostatic discharge requires isolation of all wires using insulating tape or heat shrink tubes. Metallic parts of a PC may be at hazardous voltage levels if the computer is not connected properly to the protective earth.

## 4. Microphone versus line audio input ports: where to connect the photodetector

The microphone and line audio inputs are somewhat different. As it has been shown before, the microphone input has a bias voltage and a series resistor, and it has an additional amplifier stage for small signals. This means that the input is AC-coupled to the first amplifier stage whose output is AC-coupled to the analogue-to-digital converter. This arrangement may result in a strange behaviour for photodetector signals if the gain is set high and when out-of-range signals are present. Abrupt changes in the photodetector signal will cause exponentially decaying signals after the first AC coupling stage, and if the amplification is high, the amplifier output shows clipping. This clipped signal goes into the next AC coupled stage, whose output is digitised by the analogue-to-digital converter. At this point an extra transient will appear when the output signal of the amplifier stage is no longer clipped.

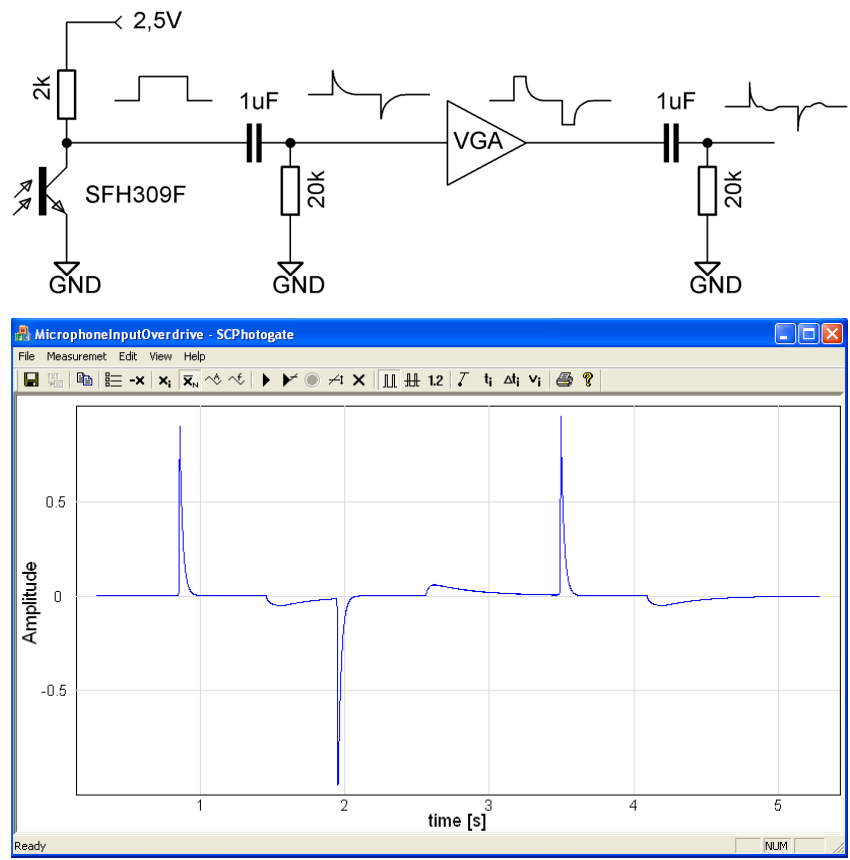

**Figure 10.** Amplitude distortion of an input signal connected to the microphone input. The parasitic transients are caused by the variable gain amplifier (VGA) of the microphone port when significant amplitude clipping occurs due to a large volume setting.

The parasitic transient can be small to start with, otherwise it can be reduced or removed by lowering the microphone gain. Note that the line input does not exhibit this behaviour due to its single AC-coupled stage; on the other hand, phototransistors and photoresistors connected to this input will require external bias as it was shown in figure 7.

There is another fact to keep in mind considering the inputs of the sound card. Automatic microphone plug-in detection and verification may need attention to set up the port properly. The software driver sometimes allows special signal processing for the microphone input. Noise reduction, echo cancellation, automatic gain can cause quite strange behaviour of the recorded waveform, therefore these features must be disabled for photodetector applications. Again, the line input does not show this behaviour.

In summary, the simplest connection of the photodetector is provided by the microphone input, but some care is needed to set up the audio features. If the detector is connected to the line input, external biasing is required, see figure 7. However, the microphone input can also be used to bias the photodetector connected to the line input as shown on figure 11.

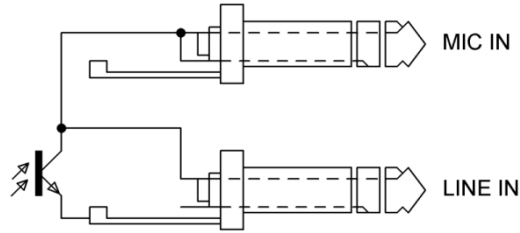

**Figure 11.** If the user prefers to use the line input, the photodetector can be biased from the microphone input.

## 5. Light source interfacing possibilities

Most of the light sources listed before are externally powered; therefore they are not interfaced to the PC. However, an infrared LED can be powered directly from the PC. The USB 5 V supply (pin 1 is 5 V, wire is red; pin 4 is GND, wire is black) easily gives enough current, but if smaller light intensity suffices, even the headphone out port can be used. Since the output signal can only be alternating, an AC-to-DC conversion is needed. One example is shown in figure12.

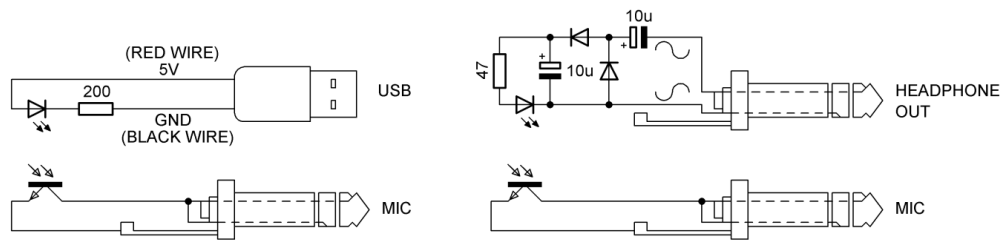

**Figure 12.** Powering the photogate LED from the computer. USB port can supply 5 V DC and a current limiting resistor of about 200 Ω provides sufficient brightness, while if smaller intensities are acceptable, even the headphone output can supply the LED after AC-to-DC conversion.

Note that even alternating powering of the LED can have advantages, as will be shown later.

## 6. Photogate operating modes

Since the input signal is AC-coupled, if the signal changes from one level to another, an exponentially decaying impulse will be observed, as it was shown in section "Protecting the audio inputs". This means that this signal does not show whether the light beam is blocked, only the changes are observable. Although this allows the detection light beam interruptions, some tricks can help to detect static light levels as well.

### 6.1. AC-mode photogates

Applying a constant light source and directly measuring the signal of a photodetector connected to the microphone input results in an AC-mode photogate, when no DC information is preserved. The time instants of the signal changing edges can be precisely detected. Several free audio recording software applications can be used to read the signal transition time instants; of these, Audacity seems to be the most popular. Although this software will do the job, its use is quite inconvenient and questionable didactically. Figure 13 shows an experiment where a flashlight served as light source, and the light beam interruptions caused by moving pendulum was detected a TEPT4400 ambient light phototransistor. Every interrupt caused an abrupt rise, and if the interrupt passed, a sharp falling edge could be observed. The microphone input volume was set to get a signal without significant amplitude clipping.

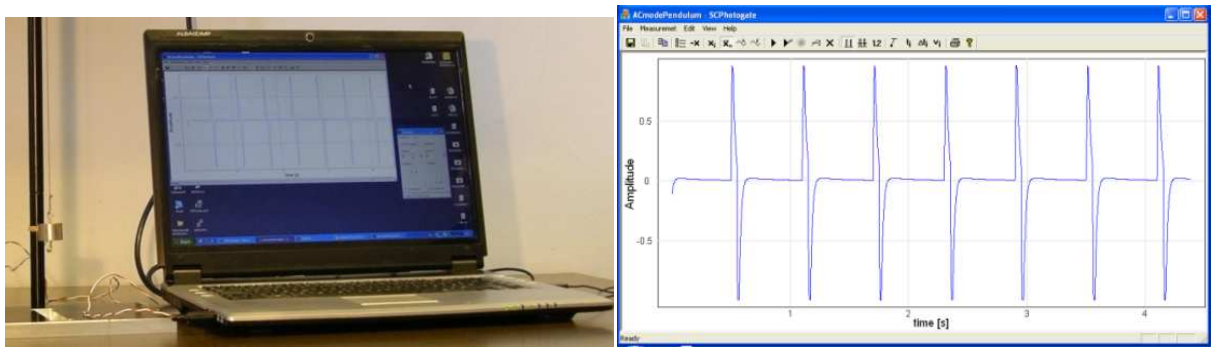

**Figure 13.** Typical signal measured with a TEPT4400 ambient light photodetector connected to the microphone input.

## 6.2. DC-mode photogates

Signal changes are not always abrupt, as they depend on several factors: the speed of the photodetector, the spatial extent of the sensing area, the light beam diameter and the shape and speed of the light interrupting object. All of these reduce the slew rate, making the detection more challenging. On the other hand, it is desirable also from a didactical point of view to implement a photogate that shows the actual state of the light beam blocking regardless of its duration. Four simple methods will be shown in the following.

### 6.2.1. Software compensation

If the signal changes are fast enough, it is quite easy to restore the light beam blocking information from the AC-coupled signal. The restored value should go high if a sharp rising edge is detected, must go low upon a falling edge, and remains unchanged otherwise. The initial state is uncertain, as one event is needed to get the correct value.

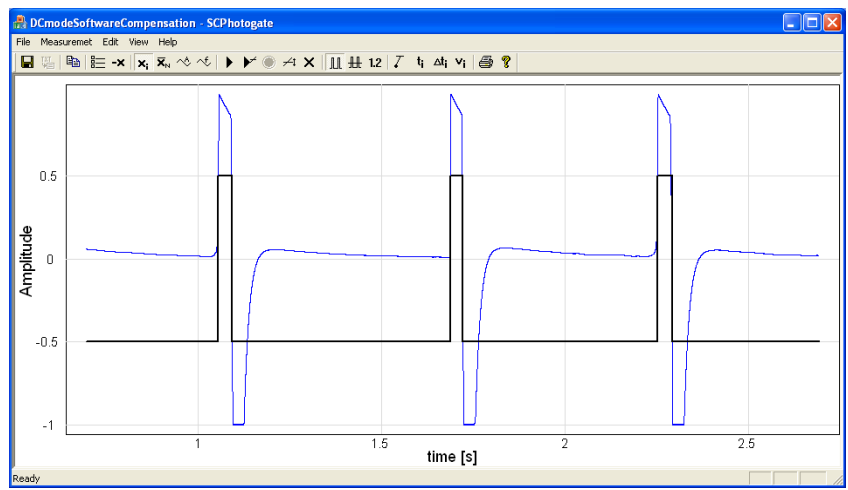

**Figure 14.** The AC coupling of the audio inputs can be compensated precisely by software if the photogate outputs sharply changing signals. The thick line shows the restored light beam blocking signal obtained using rising and falling level crossing detection values of 0.5 and -0.5, respectively.

### 6.2.2. Impedance measurement of a photoresistor

If a photoresistor is used to detect light intensity, its resistance should be measured. In the audio frequency band the impedance is almost the same as the resistance, therefore one can apply an AC voltage to drive the voltage divider formed by the photoresistor and the biasing resistor of the microphone input. The audio headphone output port can provide the required sinusoidal excitation as shown in figure 15. Note that the microphone input volume must be set to prevent amplitude clipping.

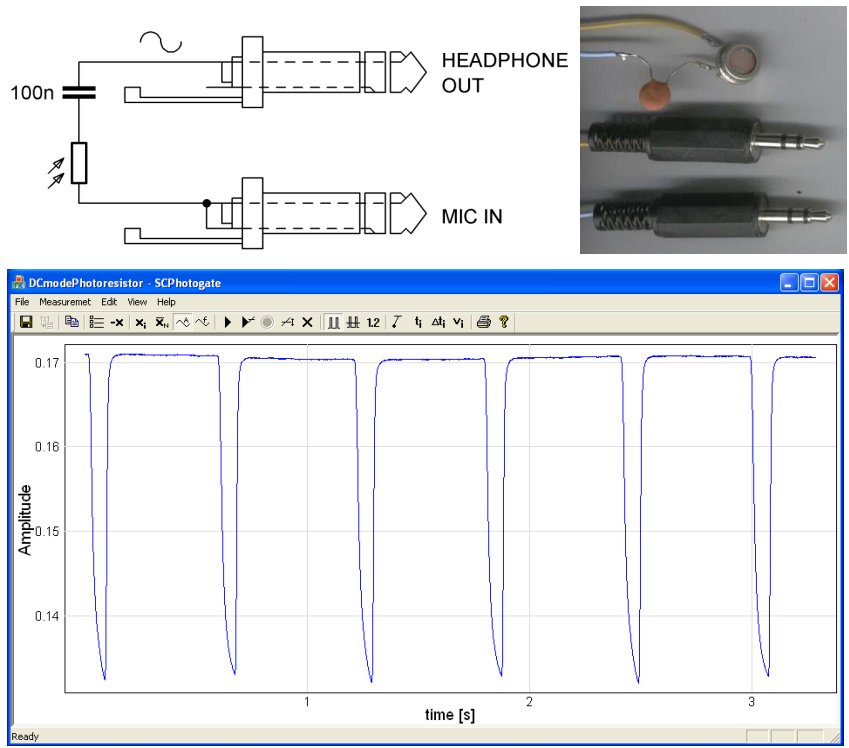

**Figure 15.** Connecting the photoresistor in impedance measurement mode. The capacitor removes any potential DC current to prevent additional transients caused by changes in light intensity.

The amplitude of the oscillating signal will correspond to the actual state of the light beam blocking. The period of excitation should be less than the desired time resolution.
One must keep in mind that here it is important to avoid overdriving the microphone input, otherwise some unwanted transients and distortion may appear in the signal.

### 6.2.3. Amplitude modulation and demodulation

Modulation-demodulation technique is a widely used tool in communications and instrumentation to encode, transfer and decode information by transforming DC or low frequency signals into a higher frequency range. It is possible to modulate the light source to get an oscillating optical signal and measure its amplitude using a phototransistor directly connected to the microphone input. Again, the amplitude of the periodic signal will reflect the actual light beam blocking state. This method has a further advantage: it attenuates all signals whose frequency is different from the excitation frequency – in other words, only the modulated source will affect the signal considerably. The time resolution is determined by the period of the oscillation. The experimental setup and results are illustrated by figure 16. The headphone output drives the LED (LD274) periodically, while the amplitude of the alternating light can be measured by a phototransistor (BPV11F) connected to the microphone input. The headphone output and microphone input volume should be set in a way to avoid amplitude saturation of the measured signal

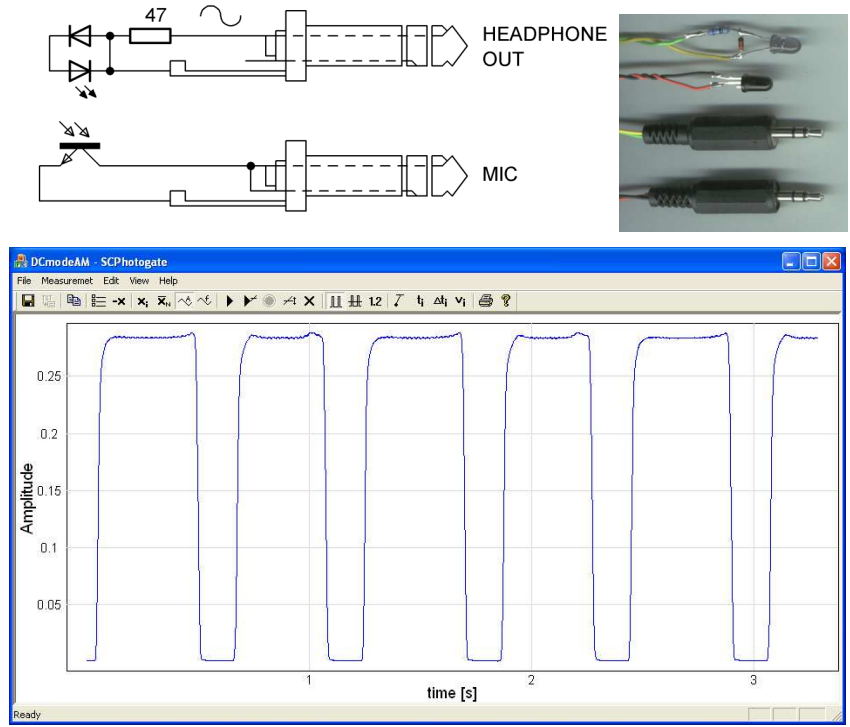

**Figure 16.** Amplitude modulation and demodulation arrangement and the typical waveform of the amplitude demodulated signal.

It is the safest to set the maximum amplitude around half of the full measurement range to prevent transients and signal distortion.

In amplitude modulation, we can apply a sine excitation whose frequency $f$ is chosen so that there is an integer number $N$ of sampling intervals $\Delta t$ in one modulation period $T$:

$$T = N\Delta t = \frac{N}{f_s}, \rightarrow \frac{f_s}{f} = N, \quad (1)$$

where $f_s$ is the sampling frequency of the sound card (typical values are $f_s = 44100$ Hz, $N = 44$ and thus $f \approx 1002.27$ Hz). This way we can avoid spectral leakage. The modulation signal is then

$$m_i = A \sin(2\pi f t_i) = A \sin\left(2\pi \frac{f_s}{N} i \Delta t\right) = A \sin\left(2\pi \frac{i}{N}\right). \quad (2)$$

In demodulation, we simply calculate the Fourier amplitude of the photogate signal $x_i$ for the last $N$ number of points, that is, for one modulation period:

$$a_j = \sqrt{\left[\frac{1}{N}\sum_{i=jN}^{(j+1)N-1} x_i \cos\left(2\pi \frac{i}{N}\right)\right]^2 + \left[\frac{1}{N}\sum_{i=jN}^{(j+1)N-1} x_i \sin\left(2\pi \frac{i}{N}\right)\right]^2}. \quad (3)$$

This amplitude $a_j$ will provide a signal proportional to light intensity at a rate $f$. Using the typical values above, the data rate is $f \approx 1002.27$ Hz, that is, the time resolution of the measurement is about 1 ms.

### 6.2.4. Frequency modulation and demodulation

If the photodetector is used in a light-to-frequency conversion circuit, one can get a DC-mode photodetector again by measuring the frequency of the output signal [14-15]. Voltage-to-frequency converters can be easily modified to implement a light-to-frequency converter. One of the simplest solutions is based on the very low-cost, popular and easy-to-use NE555 timer chip as shown on figure 17, the output of the circuit should be connected to the line input of the sound card. Alternative integrated circuits are also available (eg, AD654, XR2206, VFC32).

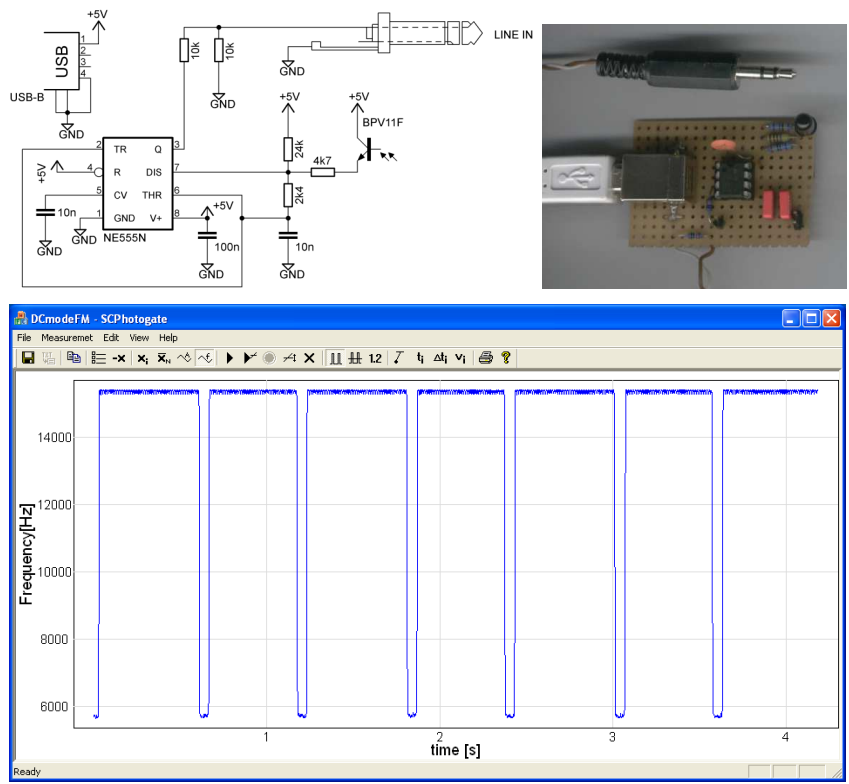

**Figure 17.** Schematic design and assembled board of the light-to-frequency converter. The instantaneous frequency of the signal clearly shows the light interruptions.

In demodulation, where the aim is to measure the instantaneous frequency, we take a fixed length of time (for instance, at a sampling rate of 44100 Hz, this can be a span of 44 data points, that is, very close to 1 ms) and simply count how many full periods are there in it. We detect the number of upward crossings $u$ and obtain the time span $t$ represented by the full periods within the measurement window (see figure 18). The average frequency within the measurement window is then $f = (u-1)/t$; this frequency will depend on light intensity. Note that the interpolation in the level crossing detection improves the resolution of the measurement.

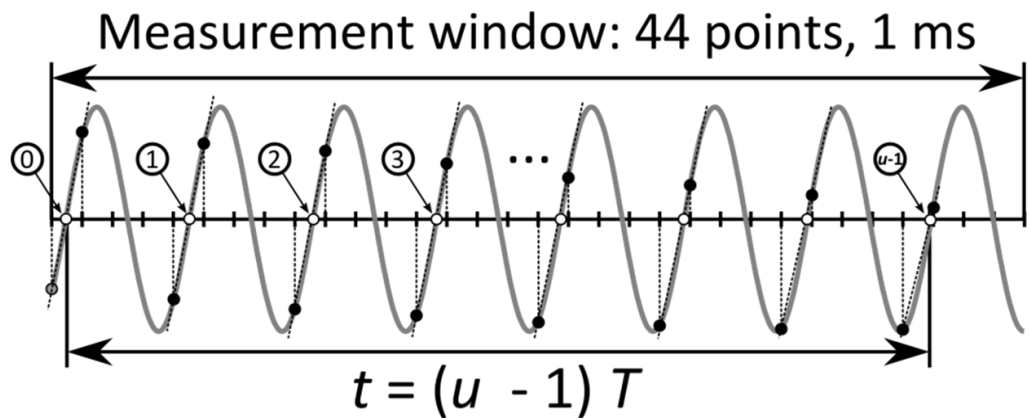

**Figure 18.** Our frequency demodulation method: we count the number of full periods in a given length of time.

Although the absolute frequency and time measurement accuracy of the sound card is much better than amplitude accuracy, the required processing is much more simple and straightforward in the case of the amplitude demodulation method. Since absolute accuracy is unimportant in photogate applications, and the frequency modulation-demodulation technique is more sensitive to noisy parasitic light signals not associated with the applied light source, the amplitude modulation/demodulation technique is preferred.

## 7. Multiple photogates

The microphone input port is monaural in most cases, only one signal can be measured at a time. There are sound cards with true stereo microphone inputs that can accept two photodetectors, as it was shown on figure 5. The line input is always stereo on all computers, but in this case one must provide the required bias either from the USB port or using a battery. Photodiodes or solar cells could be used without bias voltage, but they are not the preferred detectors.

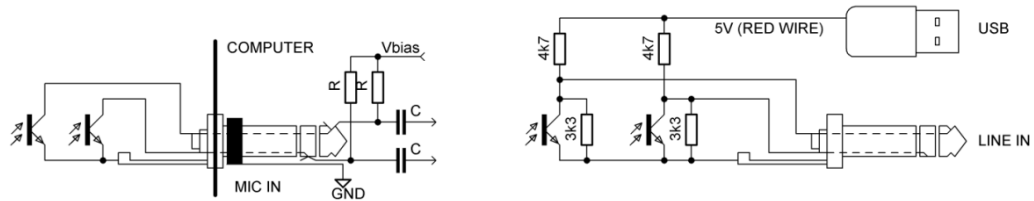

**Figure 19.** Two photodetectors can be connected to a true stereo microphone input or to the line input if they are externally powered.

It is also possible to use a single input - like the most common monaural microphone input - for multiple photodetectors, as will be shown below.

### 7.1. Non-selective connections

In most cases the state of the photogates will not be changed simultaneously, therefore it is possible to connect several phototransistors to the same input [12]. If the phototransistors are connected in a serial manner, and if any of the detectors is blocked from light, the signal will go high – in other words, the detectors operate in a logical OR fashion. If reflective photogates are used, the phototransistors must be connected in parallel to get the same behaviour.

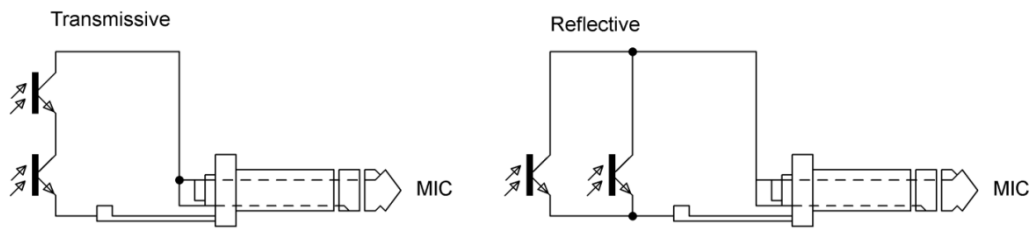

**Figure 20.** Multiple photodetectors connected to the microphone input.

Although connecting multiple phototransistors to the microphone input reduces the signal amplitude, the sound card has a high resolution (16 bits, 65536 levels), therefore the number of detectors connected to one port is quite large.

The experiment reported by figure 21 is used to detect the motion of a falling ball. A mains lamp (the small 100 Hz component is observable) illuminated the eight phototransistors from a distance of 1 m. This method is an alternative to the popular picket fence experiment to measure gravity. The user has the freedom of using different objects to compare their free fall.

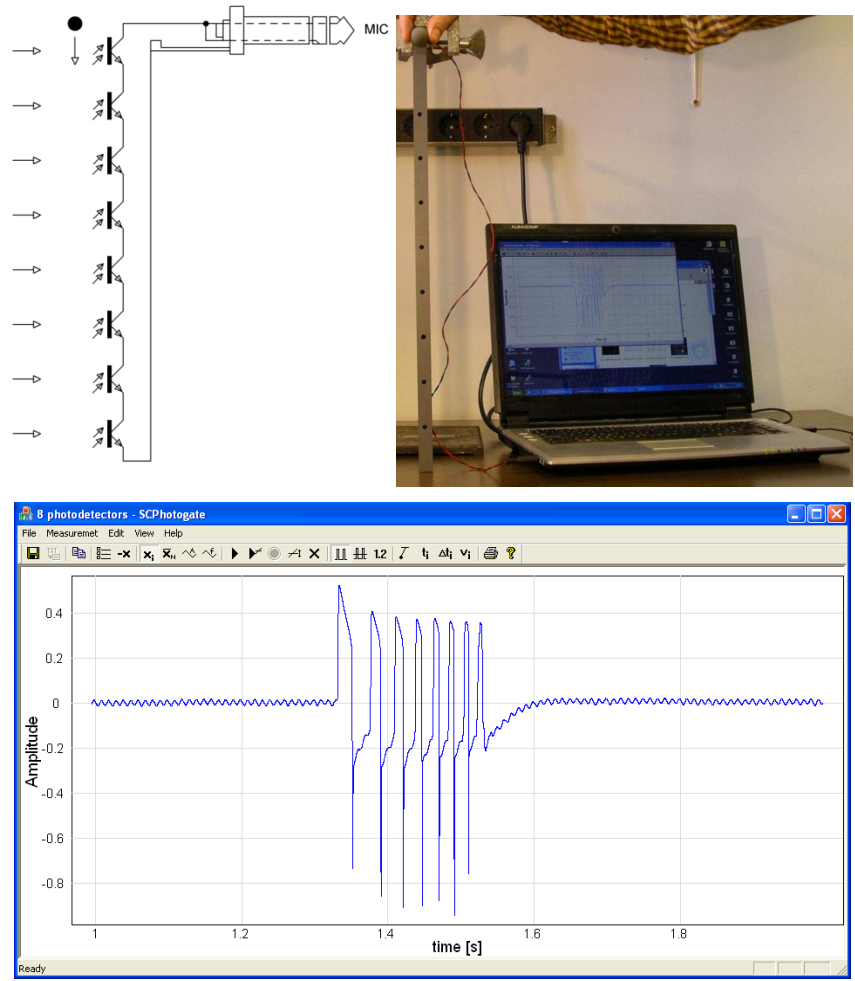

**Figure 21.** Eight BPV11F phototransistors connected directly to a microphone input can accurately detect the light interruption time instants caused by a falling ball.

## 7.2. Selective connections

In contrast to the method mentioned above, sometimes transmissive photogates are connected in parallel with different bias resistors [12]. The advantage in this case is that the signal amplitude can be used to decode which gate is blocked, however amplitude saturation must be prevented, which needs some care for AC-mode photogates, and decoding also gets more complicated. The experimental setup reported on figure 22 is used to detect the motion of a falling ball. Four phototransistors are connected in parallel to the microphone input using different weighting resistors, a mains lamp illuminated the photodetectors from a distance of 1 m. The signal amplitude depends on which detector is illuminated, therefore it is possible to identify which detector is blocked from the light beam. The microphone input volume should be set carefully to avoid amplitude saturation.

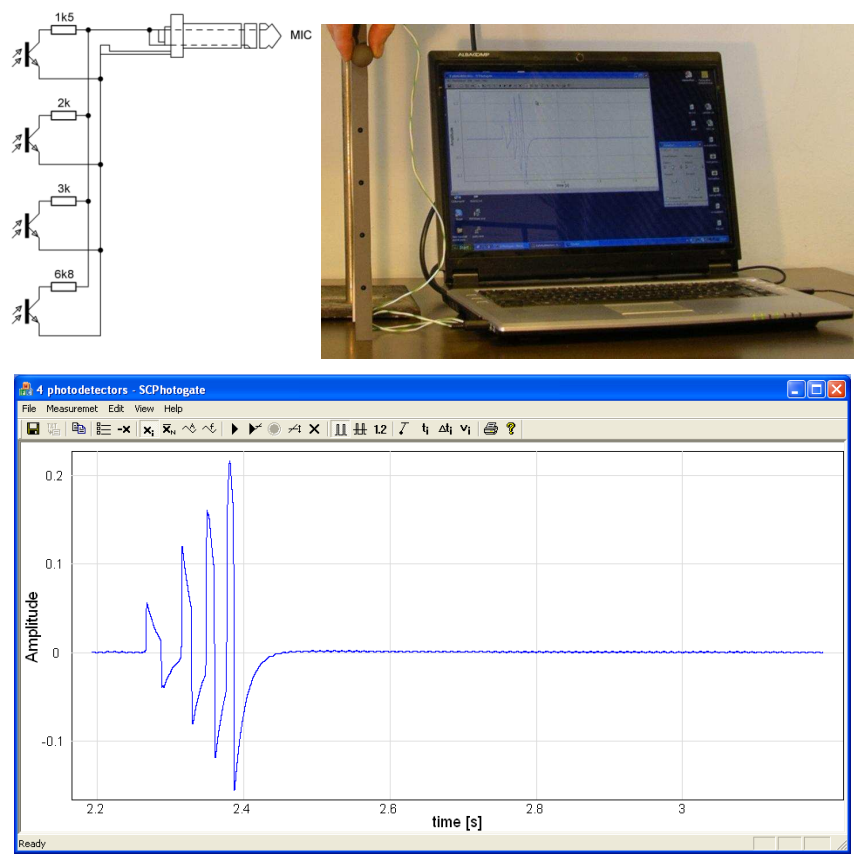

**Figure 22.** BPV11F phototransistors are connected to the microphone input using weighting resistors. The signal amplitude depends on which detector is illuminated

## 8. Software for sound card photogate applications

A special software can make sound card photogates very handy. The use of popular sound recording software is inconvenient and less transparent; getting the required information consumes considerable time, which may divert students' attention. Furthermore, the limitations make the use of all hardware realisations, like DC-mode operation, complicated.

We have developed two freely downloadable versions of software applications to make photogate experiments very easy and straightforward. One has been written in C++ and available as a single executable, can run efficiently even on slow PCs. The screenshots presented in this paper have been made using this application.

The operating principle is very similar to our open-source data acquisition software [16]. The user can start continuous data streaming, set level crossing thresholds to detect the level crossing events. Such time instants are recorded, the time between successive events are calculated and all these can be displayed versus time. Instantaneous velocity can also be derived if the length of the object is entered.

The software supports a "pendulum mode", in which the time series of the level crossings is decimated by two. In this mode, it is possible to measure the period of a pendulum's oscillations directly.

The main features of the software are summarised on figure 23. The user can set a threshold level and tolerance (hystheresis) for the photodetector signal to detect events. The time of events ($t$) and the time differences (d$t$) and the instantaneous velocity can all be displayed. An amplitude modulation-demodulation type photogate was used in pendulum mode, when only every second event is detected. The fluctuation of d$t$ is below 0.1%, and the decreasing velocity can also be clearly seen.

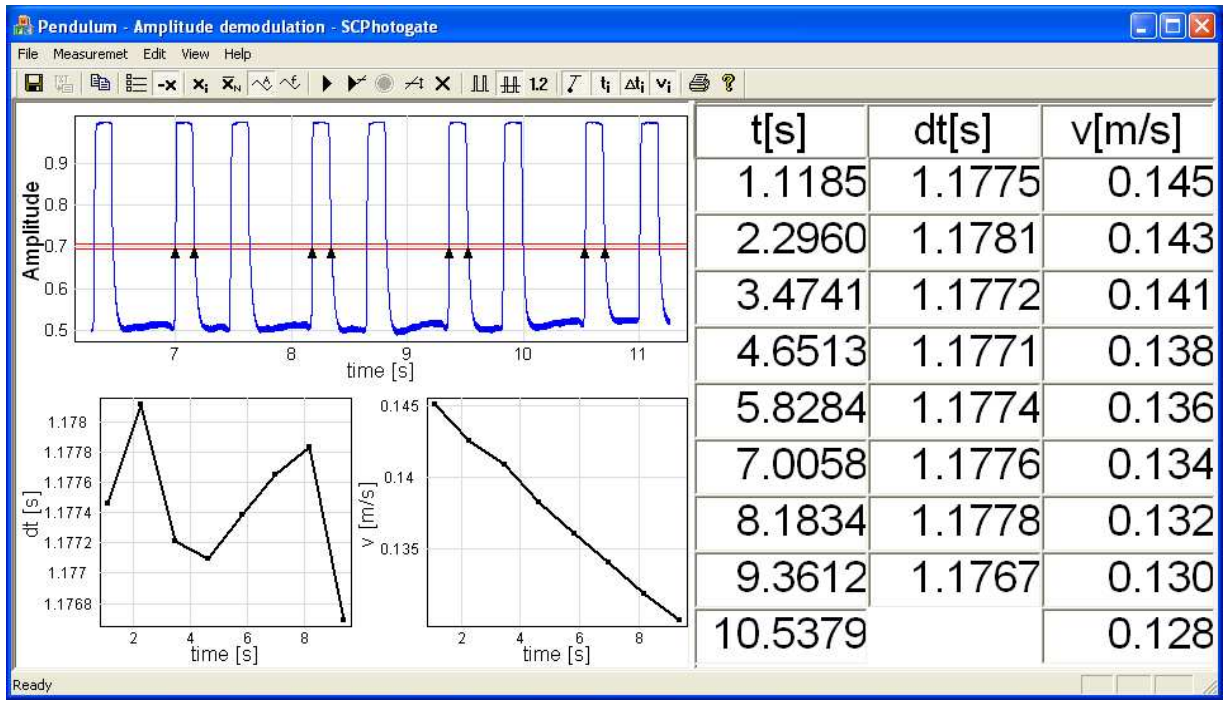

**Figure 23.** The photogate software user interface allows flexible configuration and displays comprehensive information.

DC photogate modes are also supported by software compensation, photoresistor impedance measurement and amplitude and frequency modulation-demodulation modes.

We also provide a version developed in the LabVIEW environment. This version has somewhat a reduced feature set and needs more processing power, but the source code is available, allowing experienced users to tailor the software to their needs. The LabVIEW version can easily be ported to other platforms including Linux operating systems and MacIntosh computers. Figure 24 shows the screenshot of the user interface. The user can start continuous data acquisition and set two thresholds to detect photogate signal changes. The corresponding time instants and differences are displayed in the lower part of the window. Amplitude demodulation mode is also supported by selecting the "Amplitude" tab.

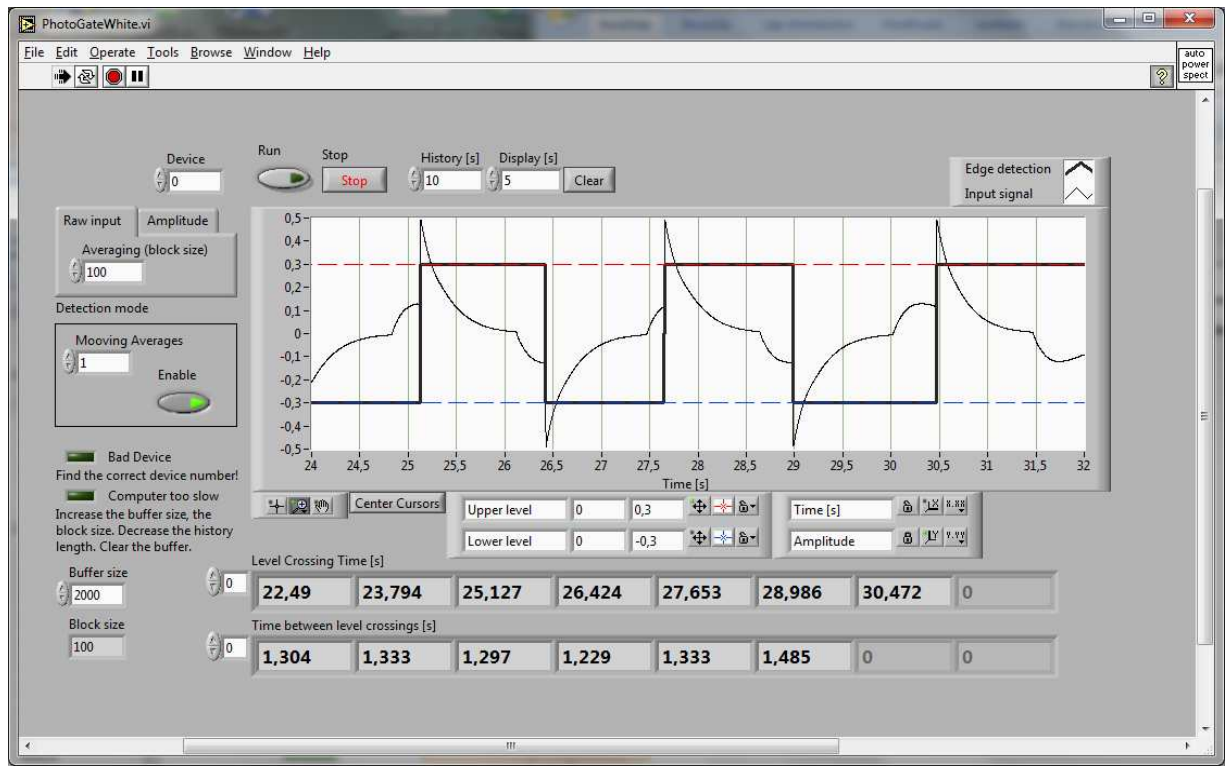

**Figure 24.** Photogate software developed in LabVIEW.

Due to the limited space available here, more detailed description and examples can be found on a dedicated web page.

## 9. Conclusions

In this paper we have reviewed the possibilities of interfacing photogates to the sound card, which is the only input on a PC to accept external analogue signals directly. Several very low cost arrangements were presented and demonstrated with experiments.

The simplest version uses a single phototransistor directly connected to the microphone input without any additional components. The cost of such a photodetector can be extremely low, about 1€.

More sophisticated arrangements were also examined and although the sound card input is AC coupled, it was shown that signal processing allows the implementation of DC-mode photogates as well at a still very impressive price of about 2€ to 5€. Due to the option of DC light measurement, these methods may offer many additional possibilities well beyond photogate applications.

An important part of our work is to present efficient software to control the photogate and analyse its output signal in real time. The software is free, especially developed for all sound card photogate arrangements and it does not need installation of any additional driver or other software. For these reasons, we believe it can be a very efficient experimentation tool – much more handy than the commonly used sound recording software.

Teachers and students can make their own photogates at incredibly low prices with very little effort. It is important to note that students can get much more inspiration, and are provided with more room for creativity, since with their own photogates they can make home exercises, homework, or just play with their computer as an efficient tool for monitoring mechanical movements or light intensity changes.

Further information can be found on a dedicated web page at http://www.noise.physx.u-szeged.hu/edudev.